\documentstyle{article}
\begin{document}
\title {
\bf \huge The Fermion-ladder Models: Extensions of the Hubbard Model with
$\eta $-pairing
}

\author{
{\bf
Heng Fan\thanks {
Permanent address: Institute of Modern Physics, 
Northwest University, Xi'an 710069, P.R.China
} }\\
\normalsize
Department of Physics, Graduate School of Science,\\
\normalsize University of Tokyo, Hongo 7-3-1,
Bunkyo-ku, Tokyo 113-0033, Japan.\\
}
\maketitle

\begin{abstract}
We propose the two-leg fermion-ladder models for $SU(2|2)$ and $SU(4)$
cases. The former is exactly the extended Hubbard model proposed by
Essler, Korepin and Schoutens. The later is a new model also with
$\eta$-pairing symmetry which is important for superconductivity. This new
extension of the Hubbard model can be solved exactly.
\end{abstract}              
       

\baselineskip 0.5truecm
Since the discovery of high-temperature superconductivity, much more
attention has been paied to the theoretical mechanism for such
phenomena. Most proposals concern about the Hubbard model and the
$t-J$ model \cite{AZR}. C.N.Yang \cite{Yang} mentioned the importance
of $\eta$-pairing mechanism and the property of off-diagonal long-range
order (ODLRO) for the eigenfunctions in superconductivity. Essler,
Korepin and Schoutens \cite{EKS} propsed an extended Hubbard model with
$\eta$-pairing symmetry. They showed that the states of
this extended Hubbard model exhibite ODLRO and is thus superconducting.

In \cite{CSAS}, the authors introduced the interlayer tunneling mechanism
to explain the superconductivity.
And the gap equation can be derived by considering two
close CuO layers described by the BCS reduced Hamiltonian and coupled
by the momentun conserving Josephson pair tunneling term.

Recently, both in experiment and theory,
there also has been growing interest in the spin ladders for
their relevance to some quasi-one-dimensional materials, which under
hole-doping may show superconductivity\cite{DR}. And many spin-ladder models
and fermion-ladder models including $t-J$ ladders and Hubbard ladders
are proposed \cite{NT}.

In this letter,
motivated by the the construction of spin-ladder models,
we study the coupled fermion models. We construct the most simple
fermion-ladder models for $SU(2|2)$ and $SU(4)$ cases.
The first one gives the extended Hubbard model which has already
been studied\cite{EKS}, the second is a new extended Hubbard model
which also has the symmetry of $\eta$-pairing and its eigenfunctions
possess ODLRO.

Generally, we will concentrate on the integrable models which can be
solved exactly. We first start from the integrable one-dimensional
fermion chain with $SU(1|1)$ symmetry. Electrons on a lattice are
described by canonical Fermi operators satisfying
$\{a^{\dagger }_i, a_j\} =\delta _{ij}$. The Fock vacuum state $|0>$
satisfies $a_i|0>=0$. The number operator for electrons on site $i$
is denoted by $n_{i,a}=a_i^{\dagger}a_i$. The Hamiltonian of $L$ site
fermion chain is written as                             
\begin{eqnarray}
H_a=\sum _{<jk>}[a_j^{\dagger }a_k+a_k^{\dagger }a_j-n_{j,a}-n_{k,a}],
\end{eqnarray}
where summation is taken for nearest neighbours.
Similarly, we suppose there is another fermion chain with different
Fermion operators $b_i^{\dagger }, b_i$ and number operator $n_{i,b}$.
And the two kind of Fermi operators still satisfy anti-commutation
relation. We thus can construct the two-leg fermion-ladder model as
\begin{eqnarray}
H_{ab}&=&t\sum _{<jk>}
[a_j^{\dagger }a_k+a_k^{\dagger }a_j-n_{j,a}-n_{k,a}
+b_j^{\dagger }b_k+b_k^{\dagger }b_j-n_{j,b}-n_{k,b}]
\nonumber \\
&&+J\sum _{<jk>}[a_j^{\dagger }a_k+a_k^{\dagger }a_j-n_{j,a}-n_{k,a}]
\times [b_j^{\dagger }b_k+b_k^{\dagger }b_j-n_{j,b}-n_{k,b}]
+U\sum _{j=1}^Ln_{j,a}n_{j,b}
\nonumber \\
&&+\mu _a\sum _{j=1}^Ln_{j,a}
+\mu _b\sum _{j=1}^Ln_{j,b}.
\end{eqnarray}
This Hamiltonian does not include all coupled terms between the two
fermion chains, but we will still reduce this Hamiltonian to a more
special case. Let $t=-1, J=-1, \mu_a=\mu_b=-U/2$, up to a constant,
this Hamiltonian is equivalent to the extended Hubbard model
proposed in \cite{EKS}. Here we use some more familiar notations,
actually we express this two kind of Fermi operators as one
kind of Fermi operators with spin down and spin up respectively,
let $a_j^{\dagger }=c^{\dagger }_{j,\uparrow }, a_j=c_{j,\uparrow },
b_j^{\dagger }=c^{\dagger }_{j,\downarrow }, b_j=c_{j,\downarrow }$,
and similarly we denote $n_{j,a}=n_{j,\uparrow },
n_{j,b}=n_{j,\downarrow }$, and also we have
$n_j=n_{j,\uparrow}+n_{j,\downarrow}$. The anticommutation relations
between Fermi operators can be expressed as
$\{c_{i,\sigma}^{\dagger}, c_{j,\tau }\} =\delta _{ij}
\delta _{\sigma \tau }$. 
We can write the fermion-ladder model as
\begin{eqnarray}
H_{eks}=-H^0_{eks}+U\sum _{j=1}^L(n_{j,\uparrow }-{1\over 2})
(n_{j,\downarrow }-{1\over 2}),
\end{eqnarray}
where the two coupled fermion chains are presented as
\begin{eqnarray}
H^0_{eks}=\sum _{<jk>}[c_{j,\uparrow}^{\dagger }c_{k,\uparrow}
+c_{k,\uparrow}^{\dagger }c_{j,\uparrow}-n_{j,\uparrow}
-n_{k,\uparrow}+1]\times 
[c_{j,\downarrow}^{\dagger }c_{k,\downarrow}+
c_{k,\downarrow}^{\dagger }c_{j,\downarrow}-n_{j,\downarrow}
-n_{k,\downarrow}+1].
\end{eqnarray}
We present the above Hamiltonian in detail as
\begin{eqnarray}
H^0_{eks}&=&\sum _{<jk>}
c_{k,\uparrow }^{\dagger }c_{j,\uparrow }
(1-n_{j,\downarrow}-n_{k,\downarrow })
+c_{j,\uparrow }^{\dagger }c_{k,\uparrow }
(1-n_{j,\downarrow}-n_{k,\downarrow })
+c_{k,\downarrow }^{\dagger }c_{j,\downarrow }
(1-n_{j,\uparrow}-n_{k,\uparrow })
\nonumber \\
&&+c_{j,\downarrow }^{\dagger }c_{k,\downarrow }
(1-n_{j,\uparrow}-n_{k,\uparrow })
+{1\over 2}(n_j-1)(n_k-1)+
c_{j,\uparrow}^{\dagger }c_{j,\downarrow}^{\dagger}      
c_{k,\downarrow}c_{k,\uparrow }
+
c_{j,\downarrow}c_{j,\uparrow }
c_{k,\uparrow}^{\dagger }c_{k,\downarrow}^{\dagger}
\nonumber \\
&&-{1\over 2}(n_{j,\uparrow }-n_{j,\downarrow})
(n_{k,\uparrow }-n_{k,\downarrow})
-c_{j,\downarrow}^{\dagger }c_{j,\uparrow}
c_{k,\uparrow}^{\dagger }c_{k,\downarrow }
-c_{j,\uparrow}^{\dagger }c_{j,\downarrow}
c_{k,\downarrow}^{\dagger }c_{k,\uparrow }
\nonumber \\
&&+(n_{j,\uparrow }-{1\over 2})(n_{j,\downarrow }-{1\over 2})
+(n_{k,\uparrow }-{1\over 2})(n_{k,\downarrow }-{1\over 2}).
\end{eqnarray}           
This Hamiltonian is just the model proposed by Essler, Korepin and
Schoutens and has the symmetry of $SU(2|2)$.
And it was studied in detail in \cite{EKS1}.

Now, let us see the $SU(4)$ case.
Generally, the $SU(2)$ fermion chain is presented as
$H_a'=\sum _{<jk>}[a_j^{\dagger }a_k+a_k^{\dagger }a_j
+2n_{j,a}n_{k,a}-n_{j,a}-n_{k,a}]$. Similar to the above method,
we can construct a two-leg fermion ladder with the $SU(4)$ symmetry
\begin{eqnarray}
H=-H^0+U\sum _{j=1}^L(n_{j,\uparrow }-{1\over 2})
(n_{j,\downarrow }-{1\over 2}),\label{Hubbard}
\end{eqnarray}
where $H^0=\sum _{<jk>}H^0_{jk}$ and we define
\begin{eqnarray}
H^0_{jk}&=&[c_{j,\uparrow}^{\dagger }c_{k,\uparrow}
+c_{k,\uparrow}^{\dagger }c_{j,\uparrow}
+2n_{j,\uparrow}n_{k,\uparrow}
-n_{j,\uparrow}
-n_{k,\uparrow}+1]
\nonumber \\
&&\times
[c_{j,\downarrow}^{\dagger }c_{k,\downarrow}+
c_{k,\downarrow}^{\dagger }c_{j,\downarrow}
+2n_{j,\downarrow}n_{k,\downarrow}
-n_{j,\downarrow}
-n_{k,\downarrow}+1].
\end{eqnarray}
We write explicitly this new Hamiltonian in the form of
$SU(4)$ generators
\begin{eqnarray}
H^0_{jk}&=&\sum _{\sigma =\uparrow ,\downarrow }
[(c_{j,\sigma}^{\dagger }c_{k,\sigma}+
c_{k,\sigma}^{\dagger }c_{j,\sigma})
(1-n_{j,-\sigma})(1-n_{k,-\sigma})
+(c_{j,\sigma}^{\dagger }c_{k,\sigma}+
c_{k,\sigma}^{\dagger }c_{j,\sigma})n_{j,-\sigma}n_{k,-\sigma}]
\nonumber \\
&&+
c_{j,\uparrow}^{\dagger }c_{j,\downarrow}^{\dagger}      
c_{k,\downarrow}c_{k,\uparrow }
+
c_{j,\downarrow}c_{j,\uparrow }
c_{k,\uparrow}^{\dagger }c_{k,\downarrow}^{\dagger}
-c_{j,\downarrow}^{\dagger }c_{j,\uparrow}
c_{k,\uparrow}^{\dagger }c_{k,\downarrow }
-c_{j,\uparrow}^{\dagger }c_{j,\downarrow}
c_{k,\downarrow}^{\dagger }c_{k,\uparrow }
\nonumber \\
&&+{1\over 2}(n_{j,\uparrow}-n_{j,\downarrow})
(n_{k,\uparrow}-n_{k,\downarrow})
+{1\over 2}(n_j-1)(n_k-1)
\nonumber \\
&&+4(n_{j,\uparrow }-{1\over 2})
(n_{j,\downarrow }-{1\over 2})(n_{k,\uparrow }-{1\over 2})
(n_{k,\downarrow }-{1\over 2})+{1\over 4}.
\end{eqnarray}
The Hamiltonian $H^0$, like the $SU(2|2)$ case,
is invariant under spin-reflection
$c_{j,\uparrow}\leftrightarrow c_{j,\downarrow }$. But unlike the
$SU(2|2)$ case, it does
not has the property of particle-hole replacement
$c^{\dagger }_{j,\sigma }\leftrightarrow c_{j,\sigma}$ invariance.
There are four kinds of state at a given site:
$|0>_i, |\uparrow >_i=c^{\dagger }_{i,\uparrow}|0>_i,
|\downarrow >_i=c^{\dagger }_{i,\downarrow}|0>_i,
|\uparrow \downarrow >_i=
c^{\dagger }_{i,\downarrow}c^{\dagger }_{i,\uparrow}|0>$, two of them
are fermionic and the other two are bosonic.  The last state
represents that an electron pair is localized on a single lattice site
which is called localons and is considered to be
the mechanism to form `Cooper pairs'\cite{EKS}.

The new Hamiltonian commute with $\eta$-pairings
$\eta ^{\dagger }
=\sum _j^Lc_{j,\downarrow}^{\dagger }
c_{j,\uparrow}^{\dagger}$ and                            
$\eta =\sum _j^Lc_{j,\uparrow}c_{j,\downarrow}$.
It is argued that $\eta$-pairs is a rather typical phenomenon in
superconductivity\cite{EKS1}. In terms of electronic operators
$c^{\dagger }_{{\bf k}\sigma}$ in momentum space,
$c_{j,\sigma }^{\dagger }={1\over {\sqrt L}}\sum _{\bf k}e^{i{\bf k}j}
c_{{\bf k}\sigma }^{\dagger }$, we find
$\eta ^{\dagger }=\sum _{\bf k}c^{\dagger }_{{\bf k}\downarrow}
c_{-{\bf k}\uparrow }^{\dagger }$
is just the BCS order parameter,
and similar for case
$\eta =\sum _{\bf k}c_{{\bf k}\uparrow}c_{-{\bf k}\downarrow }$.
                                  
The $\eta$-pairs forms a $SU(2)$ algebra.
The generators $\eta _j^{\dagger }=c_{j,\downarrow}^{\dagger }
c_{j,\uparrow}^{\dagger }$, $\eta _j=c_{j,\uparrow}c_{j,\downarrow}$ and
$\eta _j^{z}=-{1\over 2}n_j+{1\over 2}$ satisfy the relations
$[\eta _j,\eta _j^{\dagger }]=2\eta _j^z,
[\eta _j^{\dagger },\eta _j^z]=\eta _j^{\dagger },
[\eta _j,\eta _j^z]=-\eta _j$. And similarly the spin operators
$S_j=c_{j,\uparrow}^{\dagger}c_{j,\downarrow},
S_j^{\dagger}=c_{j,\downarrow}^{\dagger}c_{j,\uparrow}$ and
$S_j^z={1\over 2}(n_{j,\uparrow}-n_{j,\downarrow})$
also satisfy $SU(2)$ algebra
$[S_j,S_j^{\dagger }]=2S_j^z, [S_j^{\dagger},S_j^z]=S_j^{\dagger},
[S_j,S_j^z]=-S_j$. These generators are grassmann even (bosonic),
and generator $X_j=(n_{j,\uparrow}-{1\over 2})
(n_{j,\downarrow}-{1\over 2})$ is also a grassmann even operator.
Here we introduce eight grassmann odd (fermionic) generators
$Q_{j,\sigma }=(1-n_{j,-\sigma })c_{j,\sigma},
Q_{j,\sigma }^{\dagger }=(1-n_{j,-\sigma })c_{j,\sigma}^{\dagger},
\tilde {Q}_{j,\sigma }=n_{j,-\sigma }c_{j,\sigma},
\tilde {Q}_{j,\sigma }^{\dagger}=n_{j,-\sigma }c_{j,\sigma}^{\dagger},$
with $\sigma =\uparrow ,\downarrow $ representing
spin up and spin down respectively.
In terms of these generators, the Hamiltonian
$H^0_{jk}$ is written as
\begin{eqnarray}
H^0_{jk}&=&\sum _{\sigma =\uparrow,\downarrow}
[Q_{j,\sigma}^{\dagger}Q_{k,\sigma}+
Q_{k,\sigma}^{\dagger}Q_{j,\sigma}
+\tilde {Q}_{j,\sigma}^{\dagger}\tilde {Q}_{k,\sigma}+
\tilde {Q}_{k,\sigma}^{\dagger}\tilde {Q}_{j,\sigma}]
\nonumber \\
&&+\eta _j^{\dagger}\eta _k+\eta _j\eta _k^{\dagger }+2\eta _j^z\eta _k^z
-S_j^{\dagger }S_k-S_jS_k^{\dagger }+2S_j^zS_k^z+4X_jX_k+{1\over 4}.
\end{eqnarray}
Because we actually use the graded method, the Hamiltonian lost the
$SU(4)$ invariant property, for example, $[S^{\dagger }, H^0]\not= 0$,
with $S^{\dagger }=\sum _j^LS_j^{\dagger }$.                           
The one-dimensional Hamiltonian is equal to a graded
$SU(4)$ permutation operator
and can be diagonalized
by using the graded Bethe ansatz method. The energy of the Hamiltonian
$H^0$ is given by\cite{F}
\begin{eqnarray}
E=L-\sum _{j=1}^N\frac {1}{\lambda _j^2+{1\over 4}},
\end{eqnarray}
where $\lambda _j$ satisfy the following Bethe ansatz equations:
\begin{eqnarray}
\left( \frac {\lambda _k-{i\over 2}}{\lambda _k+{i\over 2}}\right) ^L
&=&\prod _{j=1, \not= k}^N
\frac {\lambda _k-\lambda _j-i}{\lambda _k-\lambda _j+i}
\prod _{l=1}^{N^{(1)}}\frac {\lambda _l^{(1)}-\lambda _k-{i\over 2}}
{\lambda _l^{(1)}-\lambda _k+{i\over 2}},
\\
\prod _{j=1}^{N}\frac {\lambda _k^{(1)}-\lambda _j+{i\over 2}}
{\lambda _k^{(1)}-\lambda _j-{i\over 2}}
&=&\prod _{l=1}^{N^{(1)}}\frac {\lambda _l^{(1)}-\lambda _k^{(1)}-i}
{\lambda _k^{(1)}-\lambda _l^{(1)}-i}
\prod _{j=1}^{N^{(2)}}\frac {\lambda _k^{(1)}-\lambda _j^{(2)}-{i\over 2}}
{\lambda _j^{(2)}-\lambda _k^{(1)}-{i\over 2}},
\\
\prod _{l=1,\not= k}^{N^{(2)}}\frac {\lambda _l^{(2)}
-\lambda _k^{(2)}+i}{\lambda _l^{(2)}-\lambda _k^{(2)}-i}
&=&\prod _{j=1}^{N^{(1)}}\frac {\lambda _k^{(2)}-
\lambda _j^{(1)}-{i\over 2}}
{\lambda _j^{(1)}-\lambda _k^{(2)}-{i\over 2}},
\end{eqnarray}
and $k$ take values, $1,\cdots, N$, $1\cdots N^{(1)}$ and
$1, \cdots, N^{(2)}$ respectively in the above Bethe ansatz equations.

The Hamiltonian commute with the generator $X=\sum _{j=1}^LX_j$, so
the interaction term in (\ref{Hubbard}) with coupling constant
$U$ does not break the integrability of the Hamiltonian. Generally,
We can also add such terms as chemical potential and magnetic field
in the Hamiltonian which does not change the integrability of the model.

Define the $\eta$-pairing generators $\eta =\sum _{j=1}^L\eta _j,
\eta ^{\dagger }=\sum _{j=1}^L\eta _j^{\dagger }$ and
$\eta ^z=\sum _{j=1}\eta _j^z$ which constitute the $SU(2)$
algebra\cite{EKS}.
We can find that the Hamiltonian commute with those generators
$[H,\eta ]=[H,\eta ^{\dagger }]=0$. The state
$(\eta ^\dagger )^N|0>$ is an eigenstate of the Hamiltonian
(\ref {Hubbard}) with eigenvalue $E={{UL}\over 4}-M$, here we assume that
the total number of nearest-neighbor links $<jk>$ in the lattice is $M$.
As shown in \cite{EKS}, this eigenstate possess ODLRO.
Explicitly, the off-diagonal matrix element of the reduced denstiy
matrix reads:
\begin{eqnarray}
\frac {<0|\eta ^Nc_{k,\downarrow }^{\dagger }c_{k,\uparrow}^{\dagger }
c_{l,\uparrow}c_{l,\downarrow}(\eta ^{\dagger })^N|0>}
{<0|\eta ^N(\eta ^{\dagger })^N|0>}
=\frac {N(L-N)}{L(L-1)},
\end{eqnarray}
here we assume $k\not= l$, and the relation
$<0|\eta ^N(\eta ^{\dagger })^N|0>=N!L\times \cdots \times
(L-N+1)$ is useful
in above calculation.
We can find that
the off-diagonal matrix element is constant for large distance $|k-l|$
which means that the eigenstate has the ODLRO property.
Here we showed, like the $SU(2|2)$ case, the state $(\eta )^N|0>$ which
possess ODLRO
is also an eigenstate of the new Hamiltonian (\ref{Hubbard}).

We can propose a more general extention of the Hubbard model,
besides the original terms \cite{EKS1}, we add an extra term
$W(c_{j,\sigma }^{\dagger }c_{k,\sigma }+
c_{k,\sigma }^{\dagger }c_{j,\sigma })n_{j,-\sigma }n_{k,\sigma }$
in the Hamiltonian.

The n-leg integrable fermion chains are assumed as follows
\begin{eqnarray}
H^0&=&\sum _{<jk>}\prod _{\sigma =1}^n[c_{j,\sigma }^{\dagger }
c_{k,\sigma }
+c_{k,\sigma }^{\dagger }c_{j,\sigma }-n_{j,\sigma }
-n_{k,\sigma }+1],
\\
H^0&=&\sum _{<jk>}\prod _{\sigma =1}^n[c_{j,\sigma }^{\dagger }
c_{k,\sigma }
+c_{k,\sigma }^{\dagger }c_{j,\sigma }+
2n_{j,\sigma }n_{k,\sigma }-n_{j,\sigma }
-n_{k,\sigma }+1].
\end{eqnarray}
The first one has the $SU(2^{n-1}|2^{n-1})$ symmetry, and the
second is the $SU(2^n)$ case. The Hamiltonian can
also be diagonalized by the Bethe ansatz method.

In conclusion, we propose the fermion-ladder models. For
two-leg fermion-ladder models, we obtain extentions of
the Hubbard model with $\eta$-pairing. The Hamiltonians are the
$SU(2|2)$ and $SU(4)$ cases. The $SU(4)$ case is a new
model, its eigenstate possess ODLRO property. The n-leg fermion
ladders are conjectured.

The author is supported by JSPS. He would like
to thank M.Wadati for kind support and encouragement.
He also thank A.Foerster, J.Links, Y.Z.Zhang and H.Q.Zhou
for useful communications.

\end{document}